\documentclass[journal,12pt,onecolumn,draftclsnofoot,]{IEEEtran}
\usepackage[ruled, vlined]{algorithm2e}
\usepackage{algpseudocode}
\usepackage{amsmath}
\usepackage{amssymb}
\usepackage{array}
\usepackage{arydshln}
\usepackage{blkarray, bigstrut}
\usepackage{bm}
\usepackage{booktabs}
\usepackage{braket}
\usepackage{cite}
\usepackage{enumitem}
\usepackage{graphicx}
\usepackage[mathscr]{euscript}
\usepackage{mathtools}
\usepackage{multirow}
\usepackage{framed} 
\usepackage[framed]{ntheorem}
\usepackage{nicefrac}
\usepackage{pgfplots}
\usepackage{qtree}
\usepackage{adjustbox}
\usepackage{rotating}
\pgfplotsset{compat = 1.13}
\usepackage{stfloats}
\usepackage{subfig}
\usepackage{url}
\usepackage{tikz}
\usepackage{tkz-berge}
\usetikzlibrary{patterns}
\usetikzlibrary{spy}
\usetikzlibrary{shapes, backgrounds,calc, snakes}
\usetikzlibrary{decorations.pathreplacing}
\usetikzlibrary{matrix}
\usepackage{fancybox}

\tikzstyle{vertex} = [circle, draw, inner sep = 0pt, minimum size = 10pt]
\newcommand{\vertex}{\node[vertex]}
\newframedtheorem{frm-prop}{Property}

\definecolor{bblue}{rgb}{0.11, 0.22, 0.73}
\definecolor{rred}{rgb}{1, 0.0, 0.16}
\definecolor{oorange}{rgb}{1.0, 0.55, 0.0}
\definecolor{ggreen}{rgb}{0.65, 1.0, 0.0}

\definecolor{color0}{HTML}{FF007D}
\definecolor{color1}{HTML}{760CE8}
\definecolor{color2}{HTML}{0A55FF}
\definecolor{color3}{HTML}{0DB6F4}
\definecolor{color4}{HTML}{00FF76}
\definecolor{color5}{HTML}{6FE80C}
\definecolor{color6}{HTML}{FFDE0A}
\definecolor{color7}{HTML}{FF990A}

\begin{document}

\title{Parallel and Successive Resource Allocation for V2V Communications in Overlapping Clusters}

\author{Luis F. Abanto-Leon, Arie Koppelaar, Sonia Heemstra de Groot}

\maketitle

\begin{abstract}
	The 3rd Generation Partnership Project (3GPP) has introduced in Rel. 14 a novel technology referred to as vehicle--to--vehicle (V2V) \textit{mode-3}. Under this scheme, the eNodeB assists in the resource allocation process allotting sidelink subchannels to vehicles. Thereupon, vehicles transmit their signals in a broadcast manner without the intervention of the former one. eNodeBs will thereby play a determinative role in the assignment of subchannels as they can effectively manage V2V traffic and prevent allocation conflicts. The latter is a crucial aspect to be enforced in order for the signals to be received reliably by other vehicles. To this purpose, we propose two resource allocation schemes namely bipartite graph matching-based successive allocation (BGM-SA) and bipartite graph matching-based parallel allocation (BGM-PA) which are suboptimal approaches with lesser complexity than exhaustive search. Both schemes incorporate constraints to prevent allocation conflicts from emerging. In this research, we consider overlapping clusters only, which could be formed at intersections or merging highways. We show through simulations that BGM-SA can attain near-optimal performance whereas BGM-PA is subpar but less complex. Additionally, since BGM-PA is based on inter-cluster vehicle pre-grouping, we explore different metrics that could effectively portray the overall channel conditions of pre-grouped vehicles. This is of course not optimal in terms of maximizing the system capacity---since the allocation process would be based on simplified surrogate information---but it reduces the computational complexity.
\end{abstract}

\begin{IEEEkeywords}
	weighted bipartite graph matching, radio resource allocation, broadcast vehicular communications, sidelink
\end{IEEEkeywords}

\IEEEpeerreviewmaketitle

\section{Introduction}
In the last months we have been witness to an enormous effort from academia and industry in developing novel techniques across the many fronts of vehicle--to--vehicle (V2V) communications, which is to become a pivotal role player in the fifth generation of wireless systems. Within the many use cases of V2V communications, safety-related services are unquestionably among the most important and challenging. Further enhancements capable of guaranteeing low latency and high reliability would become inestimable assets for deployment of fully-connected vehicle systems with the potential to reduce the amount of road traffic accidents \cite{b1}. Nevertheless, due to extreme mobility and highly varying channel conditions, the stringent requirements for this type of scenario are not so straightforward to fulfill \cite{b5}. Hence, V2V communications calls for further research and comprehensive field tests before it can become a trustworthy technology.
 
In this work, we consider that vehicles periodically broadcast short-term signals called cooperative awareness messages (CAMs) \cite{b2}. A CAM message---which is transported over a sidelink subchannel---contains meaningful information of a vehicle, e.g. speed, position, direction, that drivers and \slash or autonomous vehicles can harness for making improved and more rational decisions. In V2V \textit{mode-3}, a crucial target that eNodeBs must guarantee is a time-domain conflict-free assignment of subchannels \cite{b4}. Conversely to traditional cellular systems where communications are controlled by the eNodeB and are virtually point--to--point links between mobile users, in V2V \textit{mode-3} data traffic is not subject to management. For instance, if we consider a cellular system with 4 users and therefore two point--to--point links, the eNodeB can allocate the two uplink transmit users in the same time subframe but in different frequency subchannels. Afterwards, via downlink the other two users may even receive in the same subframe the corresponding data from the senders. On the other hand, V2V \textit{mode-3} operates in a broadcast manner where transmission and reception are implemented without intervention of eNodeBs. Therefore, due to the absence of a controller that dictates the uplink and downlink instants, only one vehicle in the cluster can transmit at a time while the others receive. If two or more vehicles transmit concurrently, the data sent by one will not reach the other, thus originating a conflict. Nevertheless, a subchannel that serves a vehicle in a certain cluster can be repurposed by other, if the latter vehicle belongs to a different cluster. Thus, eNodeBs will play a determinative role in effectively allocating subchannels to in-coverage vehicles.

We formulate the resource allocation problem as a weighted bipartite graph matching where the aim is to find a perfect one--to--one vertex assignment with maximal sum-rate capacity. We propose two suboptimal resource allocation approaches, namely $(i)$ bipartite graph matching-based successive allocation (BGM-SA) and $(ii)$ bipartite graph matching-based parallel allocation (BGM-PA). The former one is a cluster-wise sequential scheme that performs allocation with priority, from the most to the least constrained cluster. The latter algorithm is based on a primary stage of random vehicle pre-grouping followed by a secondary resource allocation stage. In BGM-PA, we have experimented with different metrics in order to discover one that could effectively depict the channel conditions of a set of pre-grouped vehicles, while still providing an acceptable sum-rate capacity value. We have employed the Kuhn-Munkres algorithm \cite{b6} as a basis for both algorithms. Moreover, modifications have been considered to enforce intra-cluster constraints and thus prevent conflicts. 

Our paper is structured as follows. In Section II, we explain the motivation of our work and succinctly describe our contributions. In Section III, we describe the sidelink channel structure for V2V broadcast communications. In Section IV, we formulate the resource allocation problem. In Section V and Section VI, the proposed approaches BGM-SA and BGM-PA are presented, respectively. In Section VII, we discuss simulation results in detail for several scenarios. Finally, Section VIII is devoted to summarizing our conclusions.

\section{Motivation and Contributions}
\begin{figure*}
	\begin{center}
		\begin{tikzpicture}
		\node (img) {\includegraphics[width=1\linewidth]{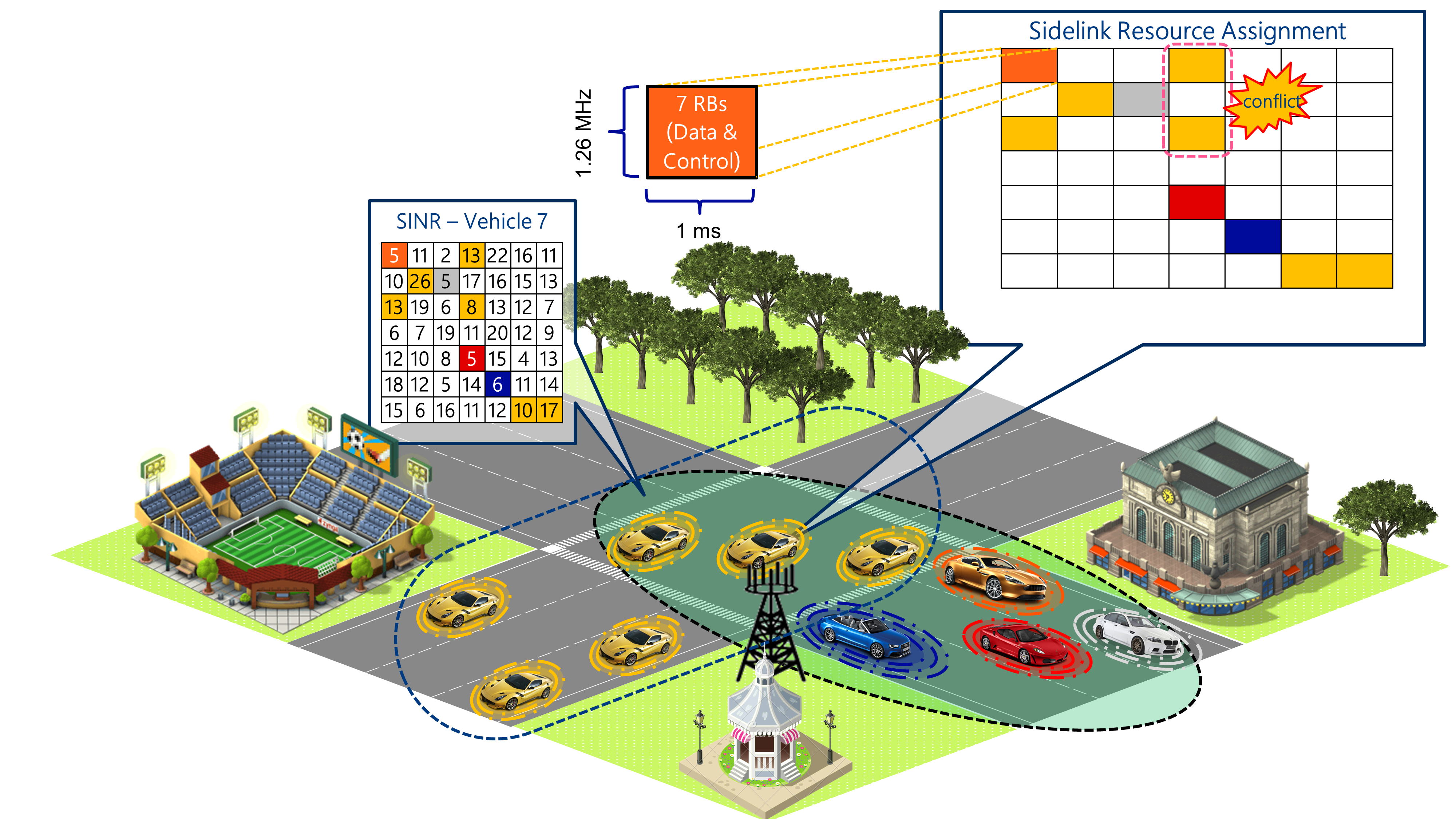}};
		\node[bblue, rotate = -18] at (5.10, -3.27) {\footnotesize \text{Vehicle $v_1$ }};
		\node[bblue, rotate = -18] at (3.50, -3.37) {\footnotesize \text{Vehicle $v_2$ }};
		\node[bblue, rotate = -18] at (3.62, -1.76) {\footnotesize \text{Vehicle $v_3$ }};
		\node[bblue, rotate = -18] at (1.55, -3.37) {\footnotesize \text{Vehicle $v_4$ }};
		\node[bblue, rotate = 22] at (1.75, -1.45) {\footnotesize \text{Vehicle $v_5$ }};
		\node[bblue, rotate = 22] at (0.4, -1.4) {\footnotesize \text{Vehicle $v_6$ }};
		\node[bblue, rotate = 22] at (-0.9, -1.25) {\footnotesize \text{Vehicle $v_7$ }};
		\node[bblue, rotate = 22] at (-1.15, -2.55) {\footnotesize \text{Vehicle $v_8$ }};
		\node[bblue, rotate = 22] at (-3.2, -1.95) {\footnotesize \text{Vehicle $v_9$ }};
		\node[bblue, rotate = 22] at (-2.75, -3.15) {\footnotesize \text{Vehicle $v_{10}$ }};
		
		\node at (5.85, 4.25) {\footnotesize $v_{10}$};
		\node at (5.85, 3.4) {\footnotesize $v_{8}$};
		\node[white] at (5.85, 2.55) {\footnotesize $v_{2}$};
		\node[white] at (6.55, 2.15) {\footnotesize $v_{4}$};
		\node at (7.25, 1.7) {\footnotesize $v_{5}$};
		\node at (7.95, 1.7) {\footnotesize $v_{7}$};
		\node at (4.45, 3.85) {\footnotesize $v_{6}$};
		\node at (5.15, 3.85) {\footnotesize $v_{1}$};
		\node at (3.75, 3.4) {\footnotesize $v_{9}$};
		\node at (3.75, 4.25) {\footnotesize $v_{3}$};
		
		\node at (3.75, 1.30) {\tiny $t = 1$};
		\node at (4.45, 1.30) {\tiny $t = 2$};
		\node at (5.15, 1.30) {\tiny $t = 3$};
		\node at (5.85, 1.30) {\tiny $t = 4$};
		\node at (6.55, 1.30) {\tiny $t = 5$};
		\node at (7.25, 1.30) {\tiny $t = 6$};
		\node at (7.95, 1.30) {\tiny $t = 7$};
	
	\end{tikzpicture}
	\caption{Sidelink V2V broadcast communications scenario}
	\label{f1}
	\end{center}
\end{figure*}

The motivation of this paper can be clearly explained through Fig. \ref{f1}. We observe two communications clusters; one consisting of 7 vehicles, namely $\{v_1, v_2, v_3, v_4, v_5, v_6, v_7\}$, whereas the remaining cluster consists of 6, i.e. $\{v_5, v_6, v_7, v_8, v_9, v_{10}\}$. While there are no conflicts in the 7-vehicle cluster---as vehicles have been assigned orthogonal time-domain subchannels---in the remaining cluster we can identify a conflict. Observe that in subframe $t = 4$, vehicles $v_8$ and $v_{10}$ have been assigned subchannels located in the same subframe. Thus, these subchannels are non-orthogonal in time domain and therefore, $v_8$ and $v_{10}$ will not be able to receive each other's information (assuming that vehicles are equipped with half-duplex PHY). In order to prevent this kind of issues from occurring, we propose two resource allocation schemes. Our contributions are summarized in the following points.

\begin{itemize}
	\item In Section IV, we introduce a compact matrix formulation for the resource allocation problem when multiple clusters are considered. 
	\item The mentioned formulation includes additional constraints to prevent intra-cluster time-domain conflicts. It also contemplates a notation for representing vehicles with multiple cluster memberships, which facilitates modeling of vehicles at intersections. 
	\item In Section V, we propose a scheme called BGM-SA which allocates subchannels to vehicles in a  sequential and hierarchical manner. BGM-SA is capable of attaining near-optimal performance at lower complexity than exhaustive search.
	\item In Section VI, we introduce a second approach called BGM-PA which is based on $(i)$ inter-cluster vehicle pre-grouping and $(ii)$ subchannel assignment. Due to pre-grouping, the performance of BGM-PA is modest compared to BGM-SA but with lower complexity.
	\item We also devise six simple metrics to optimize the allocation of subchannels in BGM-PA and the performance of each is evaluated.
\end{itemize}

\section{Sidelink Resources Channelization}
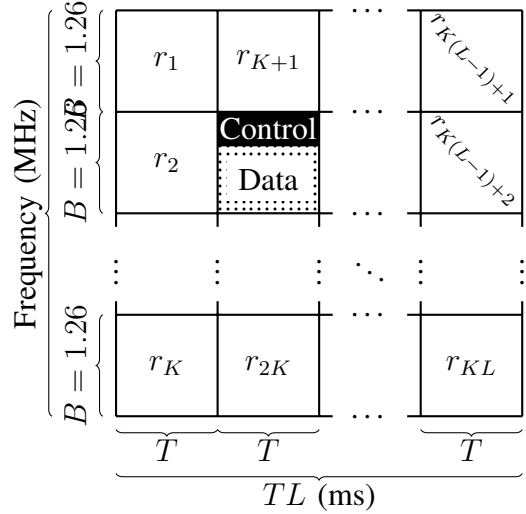
\begin{figure}[!t]
	\centering
	\begin{tikzpicture}[scale = 0.9]
	\draw[step=1.5cm, thick] (0,0) grid (3.2,-3.2);
	\draw[step=1.5cm, thick] (4.3,0) grid (6,-3.2);
	\draw[step=1.5cm, thick] (0,-4.3) grid (3.2,-6.001);
	\draw[step=1.5cm, thick] (4.3,-4.3) grid (6,-6.001);
	
	\draw[fill=black] (1.5,-1.5) rectangle (3,-2);
	\node at (2.25,-1.75) {\textcolor{white}{Control}};
	\draw[pattern=dots, pattern color=black] (1.5,-2) rectangle (3,-3);
	\node[fill = white] at (2.25,-2.5) {\textcolor{black}{Data}};
	
	\node at (3.75,0) {\dots};
	\node at (3.75,-1.5) {\dots};
	\node at (3.75,-3) {\dots};
	\node at (3.75,-4.5) {\dots};
	\node at (3.75,-6) {\dots};
	
	\node at (0, -3.75) {\vdots};
	\node at (1.5, -3.75) {\vdots};
	\node at (3, -3.75) {\vdots};
	\node at (4.5, -3.75) {\vdots};
	\node at (6, -3.75) {\vdots};
	
	\node at (3.75, -3.75) {$\ddots$};
	
	\draw[decoration={brace, raise=5pt},decorate] (1.5, -6.0) -- node[right=6pt] {} (0.0, -6.0);
	\draw[decoration={brace, raise=5pt},decorate] (3.0, -6.0) -- node[right=6pt] {} (1.5, -6.0);
	\draw[decoration={brace, raise=5pt},decorate] (6.0, -6.0) -- node[right=6pt] {} (4.5, -6.0);
	
	\draw[decoration={brace, raise=5pt},decorate] (0, -1.5) -- node[right=6pt] {} (0.0, 0);
	\draw[decoration={brace, raise=5pt},decorate] (0, -3) -- node[right=6pt] {} (0.0, -1.5);
	\draw[decoration={brace, raise=5pt},decorate] (0, -6) -- node[right=6pt] {} (0.0, -4.5);

	\draw[decoration={brace, raise=5pt},decorate] (6, -6.65) -- node[right=6pt] {} (0.0, -6.65);
	\node at (3,-7.2) {$TL$ (ms)};
	
	\draw[decoration={brace, raise=5pt},decorate] (-0.75, -6) -- node[right=6pt] {} (-0.75, 0);
	\node[rotate = 90] at (-1.3,-3) {Frequency (MHz)};
	
	\node at (0.75,-6.5) {$T$};
	\node at (2.25,-6.5) {$T$};
	\node at (5.25,-6.5) {$T$};
	
	\node at (0.75,-0.75) {$r_{1}$};
	\node at (0.75,-2.25) {$r_{2}$};
	\node at (0.75,-5.25) {$r_{K}$};
	
	\node at (2.25,-0.75) {$r_{K+1}$};
	\node at (2.25,-5.25) {$r_{2K}$};
	
	\node[rotate = -45] at (5.25,-0.75) {\small $r_{K(L-1) + 1}$};
	\node[rotate = -45] at (5.25,-2.25) {\small $r_{K(L-1) + 2}$};
	\node at (5.25,-5.25) {$r_{KL}$};
	
	\node[rotate = 90] at (-0.6,-0.75) {$B=1.26$};
	\node[rotate = 90] at (-0.6,-2.25) {$B=1.26$};
	\node[rotate = 90] at (-0.6,-5.25) {$B=1.26$};
	
	\end{tikzpicture}	
	\caption{Channelization for V2V communications}
	\label{f2}
\end{figure}

We consider that uplink/downlink and sidelink spectrum resources are decoupled from each other. We assume that the resources utilized for V2V sidelink communications are located in the intelligent transportation systems (ITS) band \cite{b3} whereas uplink/downlink spectrum resources are located in bands that usually serve cellular users. As mentioned before, in V2V \textit{mode-3} vehicles periodically broadcast CAM messages to their counterparts via sidelink \cite{b7}. However, uplink is used by vehicles to report their own channel conditions to the eNodeB. Downlink is employed for $(i)$ signaling and for $(ii)$ notifying vehicles on the subchannels they have been assigned. The channelization of sidelink spectrum resources can be regarded as a time-frequency arrangement of non-overlapping subchannels as shown in Fig. \ref{f2}. The dimensions of each subchannel are $T = 1$ ms in time and $B = 1.26$ MHz in frequency, which to the best of our understanding is sufficient for conveying a CAM message. Moreover, there are $L$ subframes and each contains $K$ subchannels. Therefore, the total number of subchannels in this formation is $KL$. Furthermore, each subchannel $r_k$ (for $k = 1, 2, \dots, KL$) consists of 7 resource blocks (RBs), where 5 RBs are used for data and 2 RBs for control.

\section{Problem Formulation}
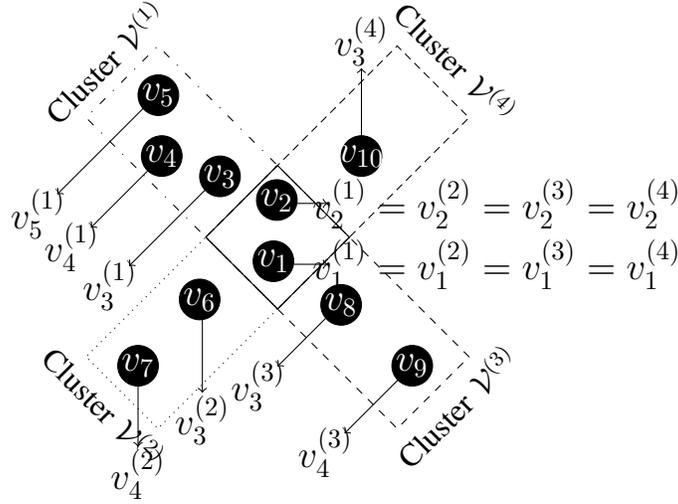
\begin{figure}[!t]
	\centering
	\[\begin{tikzpicture}[scale = 0.9]
	
	\draw[rotate around={-45:(0,0)}, dashed] (0,0) rectangle ++ (4, 1.5);
	\draw[rotate around={45:(0,0)}, densely dashed] (0, -1.5) rectangle ++ (4, 1.5);
	\draw[rotate around={225:(0,0)}, dotted] (-1.5, 0) rectangle ++ (4, 1.5);
	\draw[rotate around={135:(0,0)}, loosely dashdotted] (-1.5, -1.5) rectangle ++ (4, 1.5);
	\draw[rotate around={-45:(0,0)}, solid] (0, 0) rectangle ++ (1.5, 1.5);
	
	\draw[fill = black] (1, -0.355) circle (0.3cm) node [white] {\large $\large v_1$};
	\draw[fill = black] (1.05, 0.55) circle (0.3cm) node [white] {\large $\large v_2$};
	\draw[fill = black] (0.21, 0.9) circle (0.3cm) node [white] {\large $\large v_3$};
	\draw[fill = black] (-0.65, 1.2) circle (0.3cm) node [white] {\large $\large v_4$};
	\draw[fill = black] (-0.7, 2.1) circle (0.3cm) node [white] {\large $\large v_5$};
	\draw[fill = black] (-0.09, -0.92) circle (0.3cm) node [white] {\large $\large v_6$};
	\draw[fill = black] (-1, -1.9) circle (0.3cm) node [white] {\large $\large v_7$};
	\draw[fill = black] (2, -1) circle (0.3cm) node [white] {\large $\large v_8$};
	\draw[fill = black] (3.05, -1.9) circle (0.3cm) node [white] {\large $\large v_9$};
	\draw[fill = black] (2.3, 1.2) circle (0.3cm) node [white] {\large $\large v_{10}$};
	
	\node[rotate = 45] at (-1.5, 2.6) {Cluster $\mathcal{V}^{(1)}$};
	\node[rotate = -45] at (-1.5, -2.5) {Cluster $\mathcal{V}^{(2)}$};
	\node[rotate = 45] at (3.7, -2.5) {Cluster $\mathcal{V}^{(3)}$};
	\node[rotate = -45] at (3.7, 2.6) {Cluster $\mathcal{V}^{(4)}$};
	
	\node at (4.3, -0.4) {\large $\large v^{(1)}_1 = v^{(2)}_1 = v^{(3)}_1 = v^{(4)}_1$};
	\draw[->] (1.3, -0.4) -- (1.8, -0.4) {};
	
	\node at (4.3, 0.5) {\large $\large v^{(1)}_2 = v^{(2)}_2 = v^{(3)}_2 = v^{(4)}_2$};
	\draw[->] (1.25, 0.5) -- (1.8, 0.5) {};
	
	\draw[rotate around={-45:(0,0)}] (-0.5, -1.5) node {\large $\large v^{(1)}_3$};
	\draw[->, rotate around={-45:(0,0)}] (-0.5, 0.65) -- (-0.5, -1.1) node[rotate around={-45:(0,0)}] { };
	
	\draw[rotate around={-45:(0,0)}] (-1.3, -1.5) node {\large $\large v^{(1)}_4$};
	\draw[->, rotate around={-45:(0,0)}] (-1.3, 0.3) -- (-1.3, -1.1) node[rotate around={-45:(0,0)}] { };
	
	\draw[rotate around={-45:(0,0)}] (-2.0, -1.5) node {\large $\large v^{(1)}_5$};
	\draw[->, rotate around={-45:(0,0)}] (-2.0, 0.8) -- (-2.0, -1.1) node[rotate around={-45:(0,0)}] { };
	
	\node at (-0.05, -2.7) {\large $\large v^{(2)}_3$};
	\draw[->] (-0.05, -1.2) -- (-0.05, -2.3) {};
	
	\node at (-1, -3.5) {\large $\large v^{(2)}_4$};
	\draw[->] (-1, -2.1) -- (-1, -3.1) {};
	
	\draw[rotate around={-45:(0,0)}] (2.1, -1.0) node {\large $\large v^{(3)}_3$};
	\draw[->, rotate around={-45:(0,0)}] (2.1, 0.5) -- (2.1, -0.6) node[rotate around={-45:(0,0)}] { };
	
	\draw[rotate around={-45:(0,0)}] (3.5, -1.0) node {\large $\large v^{(3)}_4$};
	\draw[->, rotate around={-45:(0,0)}] (3.5, 0.6) -- (3.5, -0.6) node[rotate around={-45:(0,0)}] { };
	
	\node at (2.3, 2.9) {\large $\large v^{(4)}_3$};
	\draw[->] (2.3, 1.4) -- (2.3, 2.5) {};
	
	\end{tikzpicture}\]
	\caption{Overlapping vehicular clusters}
	\label{f3}
\end{figure}

Let $J$ denote the total number of partially overlapping clusters. Thus, each cluster can be denoted as a set of vehicles $\mathcal{V}^{(j)}$, each consisting of $N_j$ vehicles (for $j = 1, 2, \dots, J$). To illustrate this description, consider Fig. \ref{f3}, where the scenario is constituted by $J = 4$ partially overlapping clusters such that $\mathcal{V}^{(1)} = \{ v_1, v_2, v_3, v_4, v_5\}$, $\mathcal{V}^{(2)} = \{ v_1, v_2, v_6, v_7 \}$, $\mathcal{V}^{(3)} = \{ v_1, v_2, v_8, v_9 \}$, $\mathcal{V}^{(4)} = \{ v_1, v_2, v_{10} \}$ and cardinalities $N_1 = \lvert \mathcal{V}^{(1)} \lvert = 5, N_2 = \lvert \mathcal{V}^{(2)} \lvert = 4, N_3 = \lvert \mathcal{V}^{(3)} \lvert = 4, N_4 = \lvert \mathcal{V}^{(4)} \lvert = 3$ with vehicles $\{ v_1, v_2 \}$ lying at the intersection. Notice that each vehicle has an absolute labeling and a corresponding relative one which is with respect to the clusters a vehicle is members of \footnote{In this section only the absolute labeling is employed. The relative notation will be used in Section V, where Fig. 3 is repurposed to illustrate an example.}. In addition, there exists a set of allotable subchannels which are managed by the eNodeB. In sum, there exists a whole set of vehicles $\mathcal{V}$ distributed into $J$ clusters which are seeking to be assigned a resource from a set of allotable subchannels $\mathcal{R}$. Considering the absolute labeling, this problem can be represented as a weighted bipartite graph matching between two disjoint sets: vehicles and subchannels. Such a graph is denoted by $G(\mathcal{V}, \mathcal{R}, \mathcal{E})$, where  $\mathcal{V} = \cup_j \mathcal{V}^{(j)} = \{ v_1, v_2, \dots, v_N \}$, $\mathcal{R} = \{ r_1, r_2, \dots, r_{KL} \}$ and $\mathcal{E} = \mathcal{V} \times \mathcal{R} = \{ e_{11}, e_{12}, \dots, e_{N(KL)} \}$ is the set of edges. The total number of vehicles is denoted by $N = \sum_j \vert \mathcal{V}^{(j)} \vert - \sum_{j'}\vert \mathcal{V}^{(j)} \cap \mathcal{V}^{(j')} \vert$ for $ j \neq j'$, whereas $\hat{N} = \vert \bigcap_j \mathcal{V}^{(j)} \vert$ represents the number of vehicles at the intersection.

We can thereby represent vehicles and subchannels as vertices. Thus, the line connecting two vertices---a vehicle $v_i \in \mathcal{V}$ with a subchannel $r_k \in \mathcal{R}$---is called an edge $e_{ik}$. Each edge $e_{ik}$ has a corresponding weight $c_{ik}$ that in our case represents the achievable capacity that vehicle $v_i$ can attain in subchannel $r_k$. Therefore, $c_{ik} = B \log_2(1 + \mathsf{SINR}_{ik})$, where $B$ is the subchannel bandwidth and $\mathsf{SINR}_{ik}$ is the signal--to--interference--plus--noise ratio (SINR) that vehicle $v_i$ senses in subchannel $r_k$. The objective function is the maximization of the system sum-rate capacity subject to satisfying the allocation constraints. The two types of constraints that must be enforced are $(a)$ the intra-cluster allocation restrictions, which prevent time-domain conflicts and $(b)$ the one--to--one vertex matching conditions, which impose that each vehicle is assigned exactly one subchannel. This is equivalent to finding a vector $\bf x$ that maximizes (1a) while satisfying the constraints (1b). Thus, 
\begin{subequations} \label{e1}
	\begin{gather} 
	\begin{align}
		& {\rm max} ~ {\bf c}^T {\bf x} \\ 
		& {\rm subject~to}~
		\underbrace{
		\Bigg(
			{\left[
				\begin{array}{c}
				{\bf I}_{N \times N} \otimes {\bf 1}_{1 \times L}\\
				\hline
				{\bf Q}_{J \times N} \otimes {\bf I}_{L \times L} 
				\end{array}
			\right]} \otimes {\bf 1}_{1 \times K} \Bigg) }_{\text{constraint matrix}} {\bf x} = {\bf 1} 
	\end{align}
	\end{gather}
\end{subequations}
where $\otimes$ represents the tensor product operator, ${\bf c} \in \mathbb{R}^{M},  {\bf x} \in \mathbb{B}^M$ with $M = NLK$. ${\bf I}_{N \times N}$ and ${\bf I}_{L \times L}$ are identity matrices whereas ${\bf 1}_{1 \times L}$ and ${\bf 1}_{1 \times K}$ are vectors whose elements are all 1. ${\bf Q} \in \mathbb{B}^{J \times N}$ is the membership matrix which portrays the association of vehicles to several clusters. Thus, if a vehicle $v_i$ belongs to cluster $\mathcal{V}^{(j)}$, the element $q_{ji}$ is set to 1; otherwise it is zero. Also, $\mathbf{x} = [x_{1,1}, \dots, x_{1,KL}, \dots, x_{N,1}, \dots, x_{N,KL}]^T$, $\mathbf{c} = [c_{1,1}, \dots, c_{1,KL}, \dots, c_{N,1}, \dots, c_{N,KL}]^T$ are the solution vector and weight vector, respectively. The relation between the graph edges $e_{ik}$ and the solution vector ${\bf x}$ is the following. First, we have assumed that the graph vertices are fully connected, i.e. there are no prohibited assignments at the beginning of the resource allocation process, and therefore $e_{ik} = 1 ~\forall i, k$. The solution to the problem $\bf x$ is a subset of edges $e_{ik}$ called \textit{matching} whose weights $c_{ik}$ provide a maximal sum while respecting the constraints. Therefore, if the edge $e_{ik}$ is part of such optimal \textit{matching}, then $x_{ik} = 1$ otherwise $x_{ik} = 0$. 

\section{Proposed Algorithm BGM-SA}
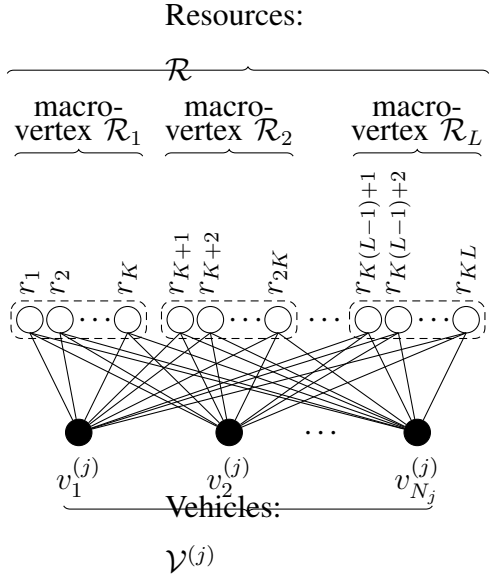
\begin{figure}[!t]
	\centering
	\[\begin{tikzpicture}][scale = 0.9]
	
	\vertex[fill] (v1) at (1.25,2.5) [label = below:$v^{(j)}_{1}$] {};
	\vertex[fill] (v2) at (3.25,2.5) [label = below:$v^{(j)}_{2}$] {};
	\node at (4.5,2.5) {\dots};
	\vertex[fill] (v3) at (5.75,2.5) [label = below:$v^{(j)}_{N_j}$] {};
	
	\vertex (r11) at (0.6,4) [label = {above:\rotatebox{90}{$r_{1}$}}] {};
	\vertex (r12) at (1,4) [label = {above:\rotatebox{90}{$r_{2}$}}] {};
	\node at (1.5,4) {\dots};
	\vertex (r13) at (1.9,4) [label = {above:\rotatebox{90}{$r_{K}$}}] {};
	
	\vertex (r21) at (2.6,4) [label = {above:\rotatebox{90}{$r_{K + 1}$}}] {};
	\vertex (r22) at (3,4) [label = {above:\rotatebox{90}{$r_{K + 2}$}}] {};
	\node at (3.5,4) {\dots};
	\vertex (r23) at (3.9,4) [label = {above:\rotatebox{90}{$r_{2K}$}}] {};
	
	\node at (4.55,4) {\dots};
	
	\vertex (r31) at (5.1,4) [label = {above:\rotatebox{90}{$r_{K(L - 1) + 1}$}}] {};
	\vertex (r32) at (5.5,4) [label = {above:\rotatebox{90}{$r_{K(L - 1) + 2}$}}] {};
	\node at (6,4) {\dots};
	\vertex (r33) at (6.4,4) [label = {above:\rotatebox{90}{$r_{KL}$}}] {};
	
	\path
	(v1) edge (0.6,3.825)
	(v1) edge (1,3.825)
	(v1) edge (1.9,3.825)
	(v1) edge (2.6,3.825)
	(v1) edge (3,3.825)
	(v1) edge (3.9,3.825)
	(v1) edge (5.1,3.825)
	(v1) edge (5.5,3.825)
	(v1) edge (6.4,3.825)
	
	(v2) edge (0.6,3.825)
	(v2) edge (1,3.825)
	(v2) edge (1.9,3.825)
	(v2) edge (2.6,3.825)
	(v2) edge (3,3.825)
	(v2) edge (3.9,3.825)
	(v2) edge (5.1,3.825)
	(v2) edge (5.5,3.825)
	(v2) edge (6.4,3.825)
	
	(v3) edge (0.6,3.825)
	(v3) edge (1,3.825)
	(v3) edge (1.9,3.825)
	(v3) edge (2.6,3.825)
	(v3) edge (3,3.825)
	(v3) edge (3.9,3.825)
	(v3) edge (5.1,3.825)
	(v3) edge (5.5,3.825)
	(v3) edge (6.4,3.825);
	
	\draw[densely dashed,rounded corners=4]($(r11)+(-.25,.25)$)rectangle($(r13)+(0.25,-.25)$);
	\draw[densely dashed,rounded corners=4]($(r21)+(-.25,.25)$)rectangle($(r23)+(0.25,-.25)$);
	\draw[densely dashed,rounded corners=4]($(r31)+(-.25,.25)$)rectangle($(r33)+(0.25,-.25)$);
	
	\draw[decoration={brace, raise=5pt},decorate] (0.4,6) -- node[right=6pt] {} (2.1,6);
	\draw[decoration={brace, raise=5pt},decorate] (2.4,6) -- node[right=6pt] {} (4.1,6);
	\draw[decoration={brace, raise=5pt},decorate] (4.9,6) -- node[right=6pt] {} (6.6,6);
	
	\node[rotate=0] at (1.25,6.8) {macro-};
	\node[rotate=0] at (3.25,6.8) {macro-};
	\node[rotate=0] at (5.75,6.8) {macro-};
	\node[rotate=0] at (1.25,6.5) {vertex $\mathcal{R}_1$};
	\node[rotate=0] at (3.25,6.5) {vertex $\mathcal{R}_2$};
	\node[rotate=0] at (5.75,6.5) {vertex $\mathcal{R}_L$};
	
	\draw[decoration={brace, raise=5pt},decorate] (0.3,7.1) -- node[right=6pt] {} (6.7,7.1);
	\draw[decoration={brace, raise=5pt},decorate] (5.95,1.7) -- node[right=6pt] {} (1.05,1.7);
	\node[text width = 2.2cm] at (3.5,1.15) {Vehicles: $\mathcal{V}^{(j)}$};
	\node[text width = 2.2cm] at (3.5,7.7) {Resources: $\mathcal{R}$};
	
	\end{tikzpicture}\]
	\caption{Constrained weighted bipartite graph}
	\label{f4}
\end{figure}

Without recurring to exhaustive search to solve (\ref{e1}), we propose to perform the allocation process in an ordered and sequential manner, which will lead to a suboptimal solution. It should be noted that, the degree of constrainedness in allocating subchannels is related to the cardinality of the cluster. Hence, the assignment of subchannels becomes more complicated when the number of vehicles in the cluster is large. Considering the foregoing facts, the allocation process in BGM-SA starts by assigning subchannels to the cluster with largest cardinality and terminates when the cluster with smallest cardinality has been processed. To illustrate this idea with an example, we consider Fig. \ref{f3}. Based on the cardinality criterion, the ordered clusters are $\lvert \mathcal{V}^{(1)} \lvert \geq \lvert \mathcal{V}^{(3)} \lvert \geq \lvert \mathcal{V}^{(2)} \lvert \geq \lvert \mathcal{V}^{(4)} \lvert$. Thus, once each of the 5 vehicles in $\mathcal{V}^{(1)}$ has been alloted a subchannel, the process will continue with cluster $\mathcal{V}^{(3)}$, where only $v_8$ and $v_9$ should be allocated since $v_1$ and $v_2$ obtained their own subchannels when $\mathcal{V}^{(1)}$ was processed. Afterwards, $v_6$ and $v_7$ will receive their respective subchannels. And the last vehicle to be serviced is $v_{10}$. At every allocation phase, vehicles must be accommodated such that they do not generate conflicts to vehicles already alloted.

To prepare the ground for the formulation of BGM-SA, we start by isolating a single cluster $\mathcal{V}^{(j)}$ as shown in Fig. \ref{f4}, where vehicles and subchannels are represented by black and white vertices, respectively. The set $\mathcal{R}$ is constituted by $KL$ vertices which are grouped into $L$ disjoint vertex subsets $\{\mathcal{R}_{l}\}_{l = 1}^L$ that we call macro-vertices, i.e. $\mathcal{R} = \cup_{l = 1}^{L} \mathcal{R}_{l}$, $\mathcal{R}_{l} \cap \mathcal{R}_{l'} = \emptyset$, $\forall l \neq l'$. Each macro-vertex $\mathcal{R}_{l}$ is a congregation of $K$ vertices, i.e. a collection of $K$ subchannels in the same time subframe. Considering the relative labeling, the bipartite graph shown in Fig. \ref{f4} is denoted by $G(\mathcal{V}^{(j)}, \mathcal{R}, \mathcal{E}^{(j)})$. Thus, the edge $e^{(j)}_{ik}$ connects vehicle $v^{(j)}_i \in \mathcal{V}^{(j)}$ with a subchannel $r_k \in \mathcal{R}$. Also, the edge weights are  defined as $c^{(j)}_{ik} = B \log_2(1 + \mathsf{SINR}^{(j)}_{ik})$. Instead of solving the allocation for the whole system in (\ref{e1}), we solve a graph matching subproblem for each cluster $\mathcal{V}^{(j)}$, for $j = 1, 2, \dots, J$. Therefore, we optimize an objective function that maximizes the sum-rate capacity of each cluster $\mathcal{V}^{(j)}$, which is expressed by 
\begin{subequations} \label{e2}
	\begin{gather} 
	\begin{align}
	& {\rm max} ~ {\bf c}^T_j {\bf x}_j \\
	& {\rm subject~to}~ 
	\underbrace{
	\Bigg(
	{\left[
		\begin{array}{c}
		{\bf I}_{N_j \times N_j} \otimes {\bf 1}_{1 \times L}\\
		\hline
		{\bf 1}_{1 \times N_j} \otimes {\bf I}_{L \times L} 
		\end{array}
		\right]} \otimes {\bf 1}_{1 \times K} \Bigg) }_{\text{constraint matrix}} {\bf x}_j = {\bf 1}
	\end{align}
	\end{gather}
\end{subequations}
where ${\bf {c}}_j \in \mathbb{R}^{M_j},  {\bf x}_j \in \mathbb{B}^{M_j}$ with $M_j = N_j K L$ and $L \geq N_j$. For completeness, we add a number of virtual vehicles with zero-valued edge weights, such that $N_j = L$ and $M_j = M = KL^2 ~ \forall j$. Therefore, the solution and weight vectors are given by $\mathbf{x}_j = [x^{(j)}_{1,1}, \dots, x^{(j)}_{1,KL}, \dots, x^{(j)}_{L,1}, \dots, x^{(j)}_{L,KL}]^T$ and $\mathbf{c}_j = [c^{(j)}_{1,1}, \dots, c^{(j)}_{1,KL}, \dots, c^{(j)}_{L,1}, \dots, c^{(j)}_{L,KL}]^T$, respectively. It is important to notice that the two types of allocation constraints mentioned in Section IV are also enforced in (2b). This means that each vehicle will be alloted exactly one subchannel and the resource allocation will guarantee that no two vehicles---in the same cluster---transmit in subchannels of same subframe. Although the constraint matrices (1b) and (2b) are similar, it is possible to exploit the structure of (2b) and further simplify the allocation problem. Recall that the time-domain orthogonality requirement on alloted subchannels is compulsory for vehicles in the same communication cluster only. It can be shown that enforcing this requirement is equivalent to aggregating vertices into macro-vertices, which in addition simplifies the complexity of (\ref{e2}), since the dimensionality is reduced. Such said vertex aggregation can be modeled as a matrix transformation, which is depicted in Fig. \ref{f5}. Thus, the problem in (\ref{e2}) can be recast as (\ref{e3})
\begin{figure}[!b]
	\centering
	\begin{tikzpicture}
	
	\draw (1,0) rectangle (3.5, -0.6) node[pos=.5] {${\bf I}_{M \times M}\otimes {\bf 1}_{1 \times K}$};
	\draw (1,-1) rectangle (3.5, -1.5) node[pos=.5] {${\bf I}_{M \times M}\otimes {\bf 1}_{1 \times K}$};
	\draw (-0.8,-1.1) rectangle (-0.5, -1.4) node[pos=.5] {$\times$};
	\draw (-2.5,-1) rectangle (-1.3, -1.5) node[pos=.5] {$diag(\cdot)$};
	
	\draw [->] (-3, -0.3) -- (1, -0.3);
	\draw [->] (-0.5, -1.25) -- (1, -1.25);
	\draw [->] (-0.65, -0.3) -- (-0.65, -1.1);
	\draw [->] (-1.3, -1.25) -- (-0.8, -1.25);
	\draw [->] (-3, -1.25) -- (-2.5, -1.25);
	\draw [->] (3.5, -0.3) -- (4, -0.3);
	\draw [->] (3.5, -1.25) -- (4, -1.25);
	
	\node[text width = 0.2cm] at (-3.3,-0.3) {${\bf x}_j$};
	\node[text width = 0.2cm] at (-3.3,-1.25) {${\bf c}_j$};
	\node[text width = 0.2cm] at (4.2,-0.3) {${\bf y}_j$};
	\node[text width = 0.2cm] at (4.2,-1.25) {${\bf d}_j$};
	
	\end{tikzpicture}
	\caption{Transformation process}
	\label{f5}
\end{figure}
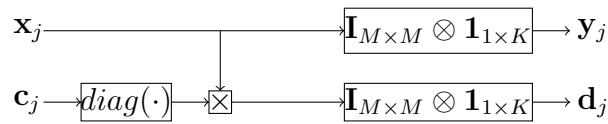
\begin{equation} \label{e3}
\hspace{-1.5cm}
\begin{array}{lclcl}
&& {\rm max} ~ {\bf d}^T_j {\bf y}_j \\
&& {\rm subject~to}~ 
{\left[
	\begin{array}{c}
	{\bf I}_{L \times L} \otimes {\bf 1}_{1 \times L}\\
	\hline
	{\bf 1}_{1 \times L} \otimes {\bf I}_{L \times L} 
	\end{array}
	\right]} {\bf y}_j = {\bf 1}
\end{array}
\end{equation}
where ${\bf y}_j = [({\bf y}_j)_{1,1}, \dots, ({\bf y}_j)_{1,L}, \dots, ({\bf y}_j)_{L,1}, \dots, ({\bf y}_j)_{L,L}]^T \in \mathbb{B}^{L^2}$ and ${\bf d}_j = \lim_{\beta \to \infty} \frac{1}{\beta} \overset{\substack{\circ}}{\log} \big\{({\bf I}_{M \times M}\otimes {\bf 1}_{1 \times K}) \mathrm{e}^{\circ \beta {\bf c}_j} \big\} \in \mathbb{R}^{L^2}$. The function $\overset{\substack{\circ}}{\log} \{\cdot\}$ represents the element-wise natural logarithm whereas $\mathrm{e}^{\circ \{ \cdot \} }$ is the Hadamard exponential \cite{b8}. Note that (\ref{e3}) is equivalent to finding a maximal matching in a graph $\widetilde{G}^{(j)} = (\mathcal{V}^{(j)}, \widetilde{\mathcal{R}}, \widetilde{\mathcal{E}}^{(j)})$ where $ \widetilde{\mathcal{R}} = \{ \tilde{r}_1, \tilde{r}_2, \dots, \tilde{r}_L \}$. Also, the edge weights between vertices in this resultant problem is ${\bf d}_j$, whose elements $d^{(j)}_{il}$ depict the weight between vertices $v^{(j)}_i$ and $\tilde{r}_l$, for $l = 1, 2, \dots, L$. Approaching (\ref{e3}) by means of finding a maximal matching in $\widetilde{G}^{(j)}$ is less complex than solving (\ref{e2}) through ${G}^{(j)}$ because $\lvert \widetilde{\mathcal{R}} \lvert$ is $K$ times smaller than $\lvert \mathcal{R} \lvert$. Thus, instead of solving either (\ref{e1}) via exhaustive search in an optimal manner or (\ref{e2}) sub-optimally through any available method, we can attain the same performance as (\ref{e2}) by solving (\ref{e3}) at lesser computational complexity.

\begin{algorithm}
	\DontPrintSemicolon
	\KwIn{A bipartite graph $\widetilde{G}^{(j)} = (\mathcal{V}^{(j)}, \widetilde{\mathcal{R}}, \widetilde{\mathcal{E}}^{(j)})$ for each cluster, such that $\big \lvert \mathcal{V}^{(j)} \big \lvert = \big \lvert \widetilde{\mathcal{R}} \big \lvert$ for completeness.} 
	\KwOut{A set of perfect matchings $\mathcal{M}^{(j)}, ~ j = 1, \dots, J$.}
	
	\Begin
	{

		\For{$j = 1: J$}
		{
			\begin{tabular}{m{0.8cm} m{5.9cm}}
		    	\underline{\textit{Step 1a:}} & Generate an initial feasible labeling $l_j$. \\
		    	\underline{\textit{Step 1b:}} & Compute the equality subgraph  $ G_l^{(j)} = $ \\ 
    								  		  & $ \{ e_{vr} \mid l_j(v) + l_j(r) = d_{vr} \} $ for $\exists v \in \mathcal{V}^{(j)}, \exists r \in \widetilde{\mathcal{R}}, e_{vr} \in \widetilde{\mathcal{E}}^{(j)}$. \\ 
		    	\underline{\textit{Step 1c:}} & Find an arbitrary matching $\mathcal{M}^{(j)}$ in $G_l^{(j)}$. \\
	 									      & \\

		    	\underline{\textit{Step 2:}} & Terminate the algorithm if the matching \\
   							     			 & $\mathcal{M}^{(j)}$ is perfect. \\
	    								 	 & \\
   
		    	\underline{\textit{Step 3:}} & Find a vertex $v' \in \mathcal{V}^{(j)}$ that has not been\\
					    	     			 & matched in $\mathcal{M}^{(j)}$ and set $\mathcal{S}^{(j)} = \{ v' \}, \mathcal{T}^{(j)} = \{ \emptyset \}$. \\ 	
		    								 & \\
					 
		    	\underline{\textit{Step 4:}} & Go to \textit{Step 6} if $N(\mathcal{S}^{(j)}) \neq \mathcal{T}^{(j)}$. \\
	    								 	 & \\
		    	
		    	\underline{\textit{Step 5a:}} & Compute the labeling $l'_j, ~ \forall$ vertex $z$ \\
    							 	  		  & \vspace{-0.3cm} $$
	    							 	 \l'_j(z)=\left\{
		    							 	 \begin{array}{ll}
			    							 	 l_j(z) - \varepsilon, ~ $if$ ~ z \in \mathcal{S}^{(j)}\\
			    							 	 l_j(z) + \varepsilon, ~ $if$ ~ z \in \mathcal{T}^{(j)}\\
			    							 	 l_j(z), ~ $otherwise$
		    							 	 \end{array}
	    							 	 \right.
	    							 	 $$ \\
	    							 	 	  & \vspace{-0.7cm} where \\
	    							 	 	  & \vspace{-1.0cm} $$\varepsilon = \min_{\substack{v \in \mathcal{S}^{(j)} \\ r \in \widetilde{\mathcal{R}} \backslash \mathcal{T}^{(j)}}} \big\{ l_j(v) + l_j(r) - d_{vr} \big\}$$ \\
	    							 	 
		   		\underline{\textit{Step 5b:}} & Compute the equality subgraph $G_l'^{(j)}$. \\
		    	\underline{\textit{Step 5c:}} & Update the equality subgraph and \\
    								  		  & labeling: $G_l^{(j)} \leftarrow G_l'^{(j)}$,  $l_j \leftarrow l'_j$. \\
    							 	  		  & \\
			 	 
				\underline{\textit{Step 6a:}} & Find a vertex $r \in N(\mathcal{S}^{(j)}) ~ \backslash ~ \mathcal{T}^{(j)}$. \\
				\underline{\textit{Step 6b:}} & Perform $\mathcal{S}^{(j)} \leftarrow \mathcal{S}^{(j)} \cup \{ u \}$, $\mathcal{T}^{(j)} \leftarrow $ \\
										 	  & $ \mathcal{T}^{(j)} \cup \{ r \}$ and go to \textit{Step 4} if $\exists e_{ur} \in \mathcal{M}^{(j)}$ such that $u \in \mathcal{V}^{(j)}$. \\
										 	  & \\
										 
				\underline{\textit{Step 7a:}} & \spaceskip  2.62em  \relax Find an alternating path \\
										 	  & $ \left\langle e_{\hat{v}_0 \hat{r}_0} \tiny{\mapsto} e_{\hat{v}_1 \hat{r}_1} \tiny{\mapsto} \dots \tiny{\mapsto} e_{\hat{v}_m \hat{r}_m} \right\rangle $ such that $\hat{v}_n \in \mathcal{V}^{(j)}, ~ \hat{r}_n \in \widetilde{\mathcal{R}}, ~ \hat{r}_m = r, ~ e_{\hat{v}_n \hat{r}_n} \in \{ G_l^{(j)} \backslash \mathcal{M}^{(j)} \}$ for $n = 0, 1, \dots , m $, $e_{\hat{v}_n \hat{r}_{n - 1}} \in \mathcal{M}^{(j)} $ for $n = 1, 2, \dots , m $. \\						 
				\underline{\textit{Step 7b:}} & Augment the previous matching $\mathcal{M}^{(j)} \leftarrow$\\
									  		  & $ \big\{ \mathcal{M}^{(j)} \cup \{ e_{\hat{v}_n \hat{r}_n} \}_{n = 0}^{n = m} \big\} \backslash \{ e_{\hat{v}_n \hat{r}_{n - 1}} \}_{n = 1}^{n = m} $ . \\
				\underline{\textit{Step 7c:}} & Go to \textit{Step 2}. \\
									  		  & \\
										 
				\underline{\textit{Step 8:}} & Update the edges in $\widetilde{\mathcal{R}}$ such that $e_{v'r'} \leftarrow$ \\ 
											 & $0, ~ \forall r' \in {\mathcal{R}}, \forall v' \in $ $\big\{ \{\mathcal{V}^{(j_{k_1})} \cap \mathcal{V}^{(j_{k_2})} \cap \dots \cap \mathcal{V}^{(j_{k_q})} \} \backslash \mathcal{V}^{(j)} \big \}$ if $e_{v'r'} \in \mathcal{M}^{(j)}$. 
		  	\end{tabular}
		}
	}
	\caption{Bipartite Graph Matching-based Successive Allocation (BGM-SA)}
	\label{a1}
\end{algorithm}

In order to solve (\ref{e3}), we propose BGM-SA which is based on \cite{b6} and shown in Algorithm \ref{a1}. Recall that since the allocation is performed in a hierarchical and sequential manner, we first sort the clusters according to their cardinality. Thus, we assume that the clusters have been labeled such that $\lvert \mathcal{V}^{(j)} \lvert \geq \lvert \mathcal{V}^{(j + 1)} \lvert$. We believe that the algorithm is self-explanatory and therefore we will not discuss the steps in detail. Instead, we introduce the following definitions in case they were necessary for its understanding. 

\vspace{0.2cm}
\underline{\textit{Labeling}:}
A feasible vertex labeling in the bipartite graph $\widetilde{G}^{(j)}$ is a real-valued function $l_j: \mathcal{V}^{(j)} \cup \widetilde{\mathcal{R}} \rightarrow \mathbb{R}$ such that $l_j(v) + l_j(r) \geq d_{vr}, ~ \forall v \in \mathcal{V}^{(j)}, ~ \forall r \in \widetilde{\mathcal{R}}$. An initial feasible labeling $l_j$ can be obtained by assigning $\displaystyle l_j(v) = \max_{\substack{r \in \widetilde{\mathcal{R}}}} d_{vr}$ and $l_j(r) = 0$. Because Algorithm \ref{a1} operates in a sequential manner processing one cluster $\widetilde{V}^{(j)}$ at a time, the $j$ indexing has been dropped to simplify the notation and thus $d_{vr}$ is equivalent to $d^{(j)}_{vr}$. \\

\vspace{-0.25cm} \underline{\textit{Equality subgraph}:}
An equality subgraph $G_l^{(j)}$ obtained from a labeling $l_j$ contains edges $e_{vr} \in \widetilde{\mathcal{E}}^{(j)}$ such that $l_j(v) + l_j(r) = d_{vr}$ holds, as described in \textit{Step 1b}. \\

\vspace{-0.25cm} \underline{\textit{Perfect matching}:}
A matching $\mathcal{M}^{(j)}$ is said to be perfect when every vertex of a graph is linked to only one edge of the matching. \\

\vspace{-0.25cm} \underline{\textit{Neighborhood of a set}:}
In a bipartite graph, the neighborhood of a vertex $v \in \mathcal{V}^{(j)}$ is defined by $N(v) = \{r \mid e_{vr} \in G_l^{(j)} \}$. Therefore, $\displaystyle N(\mathcal{S}) = \cup_{\substack{t}} N(s_t)$, $ \forall s_t \in \mathcal{S}$ (See \textit{Step 6}). \\

\vspace{-0.25cm} For each cluster $\mathcal{V}^{(j)}$, the input is a bipartite graph  $\widetilde{G}^{(j)}=(\mathcal{V}^{(j)}, \widetilde{\mathcal{R}}, \widetilde{\mathcal{E}}^{(j)})$ and the output is a matching $\mathcal{M}^{(j)}$ that will contain the association of vehicles \big(in $\mathcal{V}^{(j)}$\big) and subchannels \big(in $\widetilde{\mathcal{R}}$\big). Such matching $\mathcal{M}^{(j)}$ is a collection of edges $e^{(j)}_{il}$ that can be mapped to ${\bf y}_j$. Thus, if $e^{(j)}_{il} \in \mathcal{M}^{(j)}$, then $({\bf y}_j)_{il} = 1$ or $({\bf y}_j)_{il} = 0$ otherwise.


\section{Proposed Algorithm BGM-PA}
\begin{figure}[!t]
	\centering
	\[\begin{tikzpicture}[scale = 0.5]
	
	\draw[rotate around={-45:(0,0)}, dashed] (0,0) rectangle ++ (4, 1.5);
	\draw[rotate around={45:(0,0)}, densely dashed] (0, -1.5) rectangle ++ (4, 1.5);
	\draw[rotate around={225:(0,0)}, dotted] (-1.5, 0) rectangle ++ (4, 1.5);
	\draw[rotate around={135:(0,0)}, loosely dashdotted] (-1.5, -1.5) rectangle ++ (4, 1.5);
	\draw[rotate around={-45:(0,0)}, solid] (0, 0) rectangle ++ (1.5, 1.5);
	
	\draw[fill = black] (1, -0.355) circle (0.35cm) node [white] {\footnotesize $v_{1}$};
	\draw[fill = black] (1.05, 0.5) circle (0.35cm) node [white] {\footnotesize $v_{2}$};
	\draw[fill = black] (0.21, 0.9) circle (0.35cm) node [white] {\footnotesize $v_{3}$};
	\draw[fill = black] (-0.65, 1.2) circle (0.35cm) node [white] {\footnotesize $v_{4}$};
	\draw[fill = black] (-0.7, 2.1) circle (0.35cm) node [white] {\footnotesize $v_{5}$};
	\draw[fill = black] (-0.09, -0.92) circle (0.35cm) node [white] {\footnotesize $v_{6}$};
	\draw[fill = black] (-1, -1.9) circle (0.35cm) node [white] {\footnotesize $v_{7}$};
	\draw[fill = black] (2, -1) circle (0.35cm) node [white] {\footnotesize $v_{8}$};
	\draw[fill = black] (3.05, -1.9) circle (0.35cm) node [white] {\footnotesize $v_{9}$};
	\draw[fill = black] (2.3, 1.2) circle (0.35cm) node [white] {\footnotesize $v_{10}$};
	
	
	\begin{scope}[on background layer]
	\draw[double distance=4mm, smooth, line width=0.02cm, line join=round, line cap=round, blue, text={Positive direction}] (-0.65, 1.2)--(-1, -1.9)--(3.05, -1.9); 
	\draw[double distance=4mm, smooth, line width=0.02cm, line join=round, line cap=round, green, text={Positive direction}] (-0.7, 2.1)--(2.3, 1.2)--(2, -1)--(-0.09, -0.92); 
	\end{scope}
	
	\node[rotate = 45] at (-1.5, 2.6) {$\mathcal{V}^{(1)}$};
	\node[rotate = -45] at (-1.5, -2.5) {$\mathcal{V}^{(2)}$};
	\node[rotate = 45] at (3.7, -2.5) {$\mathcal{V}^{(3)}$};
	\node[rotate = -45] at (3.7, 2.6) {$\mathcal{V}^{(4)}$};

	\node at (3.9, -0.4) {\small $\mathcal{W}_1$};
	\draw[->, red] (1.36, -0.4) -- (3.3, -0.4) {};
	
	\node at (3.9, 0.5) {\small $\mathcal{W}_2 $};
	\draw[->, red] (1.41, 0.5) -- (3.3, 0.5) {};
	
	\node at (1.8, 3.0) {\small $\mathcal{W}_3 $};
	\draw[->, red] (1.35, 1.9)-- (1.6, 2.65) {};
	
	\node at (-2.4, -0.5) {\small $\mathcal{W}_4 $};
	\draw[->, red] (-1.25, -0.5) -- (-1.8, -0.5) {};
	
	\node at (0.25, 3.2) {\small $\mathcal{W}_5 $};
	\draw[->, red] (0.25, 1.25) -- (0.25, 2.8) {};

	\vertex[fill] (v1) at (10,2.5) [label = left:$\mathcal{W}_{1}$] {};
	\vertex[fill] (v2) at (10,1.25) [label = left:$\mathcal{W}_{2}$] {};
	\vertex[fill] (v3) at (10,0) [label = left:$\mathcal{W}_{3}$] {};
	\vertex[fill] (v4) at (10,-1.25) [label = left:$\mathcal{W}_{4}$] {};
	\vertex[fill] (v5) at (10,-2.5) [label = left:$\mathcal{W}_{5}$] {};
	\vertex[densely dashed] (r1) at (13,2.5) [label = right:$\tilde{r}_{1}$] {};
	\vertex[densely dashed] (r2) at (13,1.25) [label = right:$\tilde{r}_{2}$] {};
	\node at (13,-0.375) {\vdots};
	\vertex[densely dashed] (r5) at (13,-2.5) [label = right:$\tilde{r}_{L}$] {};
	
	\path
	(v1) edge (r1)
	(v1) edge (r2)
	(v1) edge (r5)
	
	(v2) edge (r1)
	(v2) edge (r2)
	(v2) edge (r5)
	
	(v3) edge (r1)
	(v3) edge (r2)
	(v3) edge (r5)
	
	(v4) edge (r1)
	(v4) edge (r2)
	(v4) edge (r5)
	
	(v5) edge (r1)
	(v5) edge (r2)
	(v5) edge (r5);
	
	\node[text width = 0.2cm] at (10,-3.5) {${\mathcal{W}}$};
	\node[text width = 0.2cm] at (13,-3.55) {$\widetilde{\mathcal{R}}$};
	
	\draw[->, black, ultra thick] (5.5, -0.5) -- (8.15, -0.5) {};
	\node at (6.7, 0.4) {\footnotesize Pre-};
	\node at (6.7, -0.1) {\footnotesize grouping};
	
	\draw[dashed] (1, 0) circle (4.2cm) node [white] {};
	\node at (1,-4.7) {${\mathcal{W}}$};
	
	\end{tikzpicture}\]
	\caption{Inter-cluster vehicle pre-grouping for BGM-PA}
	\label{f6}
\end{figure}
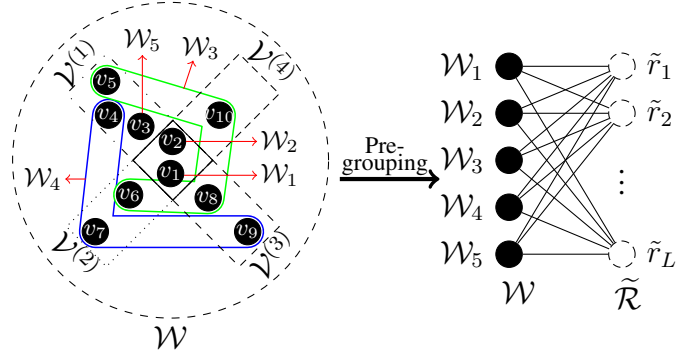
The target of inter-cluster vehicle pre-grouping in BGM-PA is to decrease the computational complexity of BGM-SA. In this regard, the allocation problem can be completed in one run by forming a virtual single cluster of vehicles, in contrast to BGM-SA that requires to allot subchannels for each cluster on a consecutive basis. Thus, each vehicle group is denoted by $\mathcal{W}_u$, such that $\mathcal{W} = \cup_u \mathcal{W}_u$ and $\mathcal{W}_u \cap \mathcal{W}_{u'} = \emptyset, ~ \forall u \neq u', u = 1, 2, \dots, \vert \mathcal{W} \lvert$. Such scenario is depicted in Fig. \ref{f6}, where the outcome of pre-grouping is shown. $\widetilde{\mathcal{R}}$ is the set of subchannels whereas $\tilde{r}_l ~ (l = 1, 2, \dots, L)$ are the same resources we referred to in (\ref{e3}). Hence, there are $\vert \mathcal{W} \lvert = 5$ groups of vehicles: $\mathcal{W}_1 = \{v_1\}$, $\mathcal{W}_2 = \{v_2\}$, $\mathcal{W}_3 = \{ v_5, v_6, v_8, v_{10} \}$, $\mathcal{W}_4 = \{ v_4, v_7, v_9 \}$, $\mathcal{W}_5 = \{v_3\}$. Note that grouping is applied only to those vehicles that do not lie at the intersection. The selection of vehicles per cluster is done randomly but aiming at assembling as many vehicles as possible. For instance, $\mathcal{W}_3$ contains 4 vehicles because that was the maximum number allowable, i.e. one vehicle per each cluster at most. On the other hand, $\mathcal{W}_4$ consists of 3 vehicles. Finally, $v_3$ was the last vehicle remaining and therefore, it by itself constitutes $\mathcal{W}_5$. Although pre-grouping is beneficial for decreasing the allocation complexity, it also originates difficulties on how to represent the overall channel conditions of each collection of vehicles $\mathcal{W}_u$. The formulation of this problem is similar to (\ref{e2}), except that $J = 1$, because after pre-grouping there will exist one cluster only. Therefore, the problem can be further reduced and thus adopt a form identical to (\ref{e3}). Nevertheless, instead of employing Algorithm \ref{a1}, we use Algorithm \ref{a2}, which in essence is similar. A central issue to take into consideration is that the resultant edge weights $\tilde{d}_{ul}$ between $\mathcal{W}_u$ and $\tilde{r}_l$, must be a joint metric that can fairly represent the overall channel conditions of a group of vehicles. Therefore, if such a group is defined by $\mathcal{W}_u = \{ w_{u1}, w_{u2}, \dots, w_{u(m_u)} \}$ with $m_u$ representing the number of vehicles in the group, then the resultant edge weight is $\tilde{d}_{ul} = metric(d_{(u1)l}, d_{(u2)l}, \dots, d_{(um_u)l})$. To this purpose, we have devised six different metrics that are defined as follows.\\

\vspace{-0.2cm}\underline{\textit{Minimum (MIN)}:} For each $\tilde{r}_l$, select the smallest edge weight $\tilde{d}_{ul}$ among all the vehicles $v' \in \mathcal{W}_u$. This is a plausible metric because if fair allocation can be guaranteed for the least favored vehicle, then the other vehicles in the group will at least experience equal or better channel conditions.\\

\vspace{-0.2cm} \underline{\textit{Maximum (MAX)}:} This metric is similar to the previous one, except that the maximum value is chosen instead of the minimum. \\

\vspace{-0.2cm} \underline{\textit{Average (AVE)}:} This metric considers the average channel conditions of all the vehicles in the group.\\

\vspace{-0.2cm} \underline{\textit{Inverse of variance (IVAR)}:} This metric measures the deviation of the channel conditions in a group of vehicles. If \textit{IVAR} is large, then the channel conditions span a large range of qualities. When \textit{IVAR} is small, we can only infer that the channel conditions are similar for the vehicles but it is difficult to know whether these are good or not.\\

\vspace{-0.2cm} \underline{\textit{Minimum plus maximum (MPM)}:} This is a merged metric that considers the overall effect of \textit{MIN} and \textit{MAX} metrics. \\

\vspace{-0.2cm} \underline{\textit{Combined metrics (COMB)}:} This metric combines some of the metrics described above. Specifically to overcome the shortcoming of \textit{IVAR} and exploit the reasoning behind \textit{MIN}, we define \textit{COMB} = \textit{AVE} + \textit{MIN} - $\sqrt{\textit{VAR}}$, where \textit{VAR} denotes variance. 

The computational complexity of exhaustive search is $\mathcal{O}({\vert \mathcal{R} \vert! / (\vert \mathcal{R} \vert - \vert \mathcal{V} \vert})!)$. On the other hand, when BGM-SA is solved through Algorithm \ref{a1} after dimensionality reduction via (\ref{e3}), the complexity is $\mathcal{O}(\max \{ J \vert \mathcal{V} \vert, J \vert \widetilde{\mathcal{R}} \vert \}^3) = \mathcal{O}(\max \{ J \vert \mathcal{V} \vert, \frac{J}{K} \vert \mathcal{R} \vert \}^3)$ whereas the complexity of BGM-PA is $\mathcal{O}(\max \{ \vert \mathcal{V} \vert, \frac{1}{K}\vert \mathcal{R} \vert \}^3)$.

\begin{algorithm}[!t]
	\DontPrintSemicolon
	\KwIn{A bipartite graph $G=(\mathcal{W}, \widetilde{\mathcal{R}}, \mathcal{E})$.} 
	\KwOut{A perfect matching $\mathcal{M}$.}
	
	\Begin
	{
		Drop the index $j$ from Algorithm 1. \\
		Perform random pre-grouping of vehicles. \\
		Select an edge metric for the grouped sets of vehicles. \\
		Perform from \textit{Step 1} to \textit{Step 7}.
	}
	\caption{Bipartite Graph Matching-based Parallel Allocation (BGM-PA)}
	\label{a2}
\end{algorithm}

\section{Simulations}
In this section, we experiment with several configurations considering different number of clusters and vehicles. We also vary the number of vehicles at the intersection in order to understand the impact on the allocation performance. We evaluate exhaustive search, BGM-SA and BGM-PA using its six variants. In our system, we consider a message rate of 10 Hz and therefore, a new allocation is performed every 100 ms for all the vehicles. In all the experiments shown onwards, we have averaged the results over 1000 simulations. In Fig. \ref{f7}, we have considered $J = 3$ clusters with $N_1 = 100$, $N_2 = 90$ and $N_3 = 80$ vehicles. The number of vehicles at the intersection is $\hat{N} = 30$ whereas the amount of vehicles in the system is $N = 210$. We have also chosen $K = 7$ and $L = 100$. 
\begin{figure*}[!t]
	\centering
	\begin{tikzpicture}
	\begin{axis}[
	ybar,
	ymin = 0,
	ymax = 13,
	width = 18.5cm,
	height = 5cm,
	bar width = 9pt,
	tick align = inside,
	x label style={align=center, font=\footnotesize,},
	ylabel = {Rate [Mbits / s / subchannel]},
	y label style={at={(-0.025,0.5)}, font=\footnotesize,},
	nodes near coords,
	every node near coord/.append style={font=\fontsize{6}{5}\selectfont, rotate = 90, anchor = west},
	nodes near coords align = {vertical},
	symbolic x coords = {Highest-Rate Vehicle, System Average Rate, Worst-Rate Vehicle, Second Worst-Rate Vehicle, System Rate Standard Deviation},
	x tick label style = {text width = 3.5cm, align = center, font = \footnotesize,},
	xtick = data,
	enlarge y limits = {value = 0.3, upper},
	enlarge x limits = 0.13,
	legend columns=4,
	legend pos = north east,
	legend style={font=\fontsize{8}{7}\selectfont, text width=3.45cm,text height=0.02cm,text depth=.ex, fill = none, }]
	]
	\addplot[fill = color0] coordinates {(Highest-Rate Vehicle,  8.97) (System Average Rate, 8.19) (Worst-Rate Vehicle, 6.89) (Second Worst-Rate Vehicle, 7.03) (System Rate Standard Deviation, 0.51)}; \addlegendentry{Exhaustive Search}
	
	\addplot[fill = color1] coordinates {(Highest-Rate Vehicle,  8.97) (System Average Rate, 8.16) (Worst-Rate Vehicle, 6.88) (Second Worst-Rate Vehicle, 7.01) (System Rate Standard Deviation, 0.51)}; \addlegendentry{Proposed BGM-SA}
	
	\addplot[fill = color2] coordinates {(Highest-Rate Vehicle,  8.97) (System Average Rate, 6.64) (Worst-Rate Vehicle, 5.32) (Second Worst-Rate Vehicle, 5.38) (System Rate Standard Deviation, 0.86)}; \addlegendentry{Proposed BGM-PA-MIN}
	
	\addplot[fill = color3] coordinates {(Highest-Rate Vehicle,  8.97) (System Average Rate, 6.19) (Worst-Rate Vehicle, 1.65) (Second Worst-Rate Vehicle, 1.95) (System Rate Standard Deviation, 2.17)}; \addlegendentry{Proposed BGM-PA-MAX}
	
	\addplot[fill = color4] coordinates {(Highest-Rate Vehicle,  8.97) (System Average Rate, 6.86) (Worst-Rate Vehicle, 3.81) (Second Worst-Rate Vehicle, 4.12) (System Rate Standard Deviation, 1.13)}; \addlegendentry{Proposed BGM-PA-AVE}
	
	\addplot[fill = color5] coordinates {(Highest-Rate Vehicle,  8.97) (System Average Rate, 5.02) (Worst-Rate Vehicle, 1.54) (Second Worst-Rate Vehicle, 1.82) (System Rate Standard Deviation, 1.71)}; \addlegendentry{Proposed BGM-PA-IVAR}

	\addplot[fill = color6] coordinates {(Highest-Rate Vehicle,  8.97) (System Average Rate, 6.79) (Worst-Rate Vehicle, 4.11) (Second Worst-Rate Vehicle, 4.31) (System Rate Standard Deviation, 1.23)}; \addlegendentry{Proposed BGM-PA-MPM}
	
	\addplot[fill = color7] coordinates {(Highest-Rate Vehicle,  8.97) (System Average Rate, 6.65) (Worst-Rate Vehicle, 5.37) (Second Worst-Rate Vehicle, 5.44) (System Rate Standard Deviation, 0.81)}; \addlegendentry{Proposed BGM-PA-COMB}
	
	\end{axis}
	\end{tikzpicture}
	\caption{Data rate for $N = 210, L = 100$ and $K = 7$ with $J=3, N_1=100, N_2=90, N_3=80, \hat{N} = 30$}
	\label{f7}
	\vspace{-0.205cm}
\end{figure*}
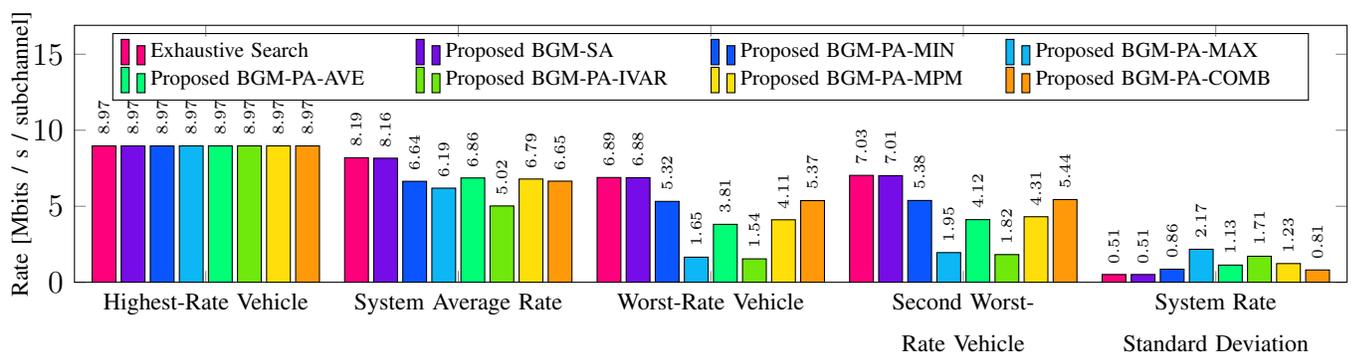
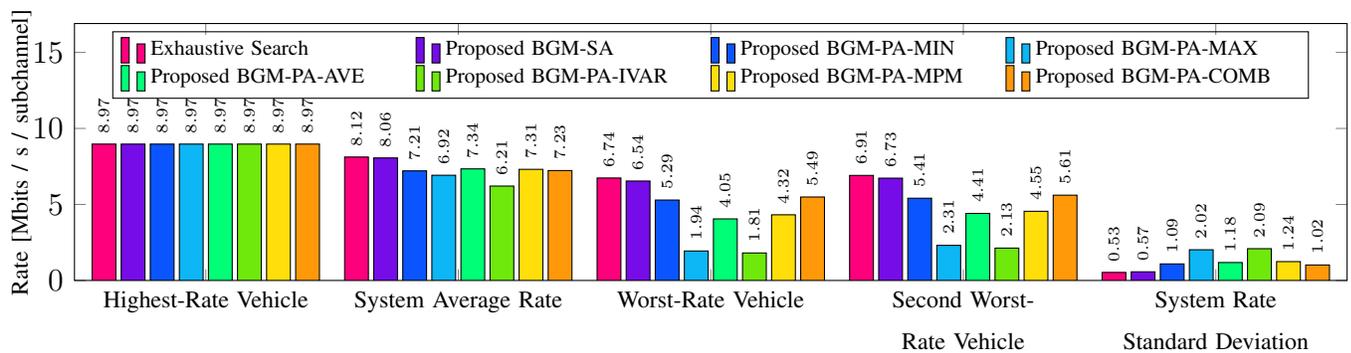
\begin{figure*}[!t]
	\centering
	\begin{tikzpicture}
	\begin{axis}[
	ybar,
	ymin = 0,
	ymax = 13,
	width = 18.5cm,
	height = 5cm,
	bar width = 9pt,
	tick align = inside,
	x label style={align=center, font=\footnotesize,},
	ylabel = {Rate [Mbits / s / subchannel]},
	y label style={at={(-0.025,0.5)}, font=\footnotesize,},
	nodes near coords,
	every node near coord/.append style={font=\fontsize{6}{5}\selectfont, rotate = 90, anchor = west},
	nodes near coords align = {vertical},
	symbolic x coords = {Highest-Rate Vehicle, System Average Rate, Worst-Rate Vehicle, Second Worst-Rate Vehicle, System Rate Standard Deviation},
	x tick label style = {text width = 3.5cm, align = center, font = \footnotesize,},
	xtick = data,
	enlarge y limits = {value = 0.3, upper},
	enlarge x limits = 0.13,
	legend columns=4,
	legend pos = north east,
	legend style={font=\fontsize{8}{7}\selectfont, text width=3.45cm,text height=0.02cm,text depth=.ex, fill = none, }]
	]
	\addplot[fill = color0] coordinates {(Highest-Rate Vehicle,  8.97) (System Average Rate, 8.12) (Worst-Rate Vehicle, 6.74) (Second Worst-Rate Vehicle, 6.91) (System Rate Standard Deviation, 0.53)}; \addlegendentry{Exhaustive Search}
	
	\addplot[fill = color1] coordinates {(Highest-Rate Vehicle,  8.97) (System Average Rate, 8.06) (Worst-Rate Vehicle, 6.54) (Second Worst-Rate Vehicle, 6.73) (System Rate Standard Deviation, 0.57)}; \addlegendentry{Proposed BGM-SA}
	
	\addplot[fill = color2] coordinates {(Highest-Rate Vehicle,  8.97) (System Average Rate, 7.21) (Worst-Rate Vehicle, 5.29) (Second Worst-Rate Vehicle, 5.41) (System Rate Standard Deviation, 1.09)}; \addlegendentry{Proposed BGM-PA-MIN}
	
	\addplot[fill = color3] coordinates {(Highest-Rate Vehicle,  8.97) (System Average Rate, 6.92) (Worst-Rate Vehicle, 1.94) (Second Worst-Rate Vehicle, 2.31) (System Rate Standard Deviation, 2.02)}; \addlegendentry{Proposed BGM-PA-MAX}
	
	\addplot[fill = color4] coordinates {(Highest-Rate Vehicle,  8.97) (System Average Rate, 7.34) (Worst-Rate Vehicle, 4.05) (Second Worst-Rate Vehicle, 4.41) (System Rate Standard Deviation, 1.18)}; \addlegendentry{Proposed BGM-PA-AVE}
	
	\addplot[fill = color5] coordinates {(Highest-Rate Vehicle,  8.97) (System Average Rate, 6.21) (Worst-Rate Vehicle, 1.81) (Second Worst-Rate Vehicle, 2.13) (System Rate Standard Deviation, 2.09)}; \addlegendentry{Proposed BGM-PA-IVAR}
	
	\addplot[fill = color6] coordinates {(Highest-Rate Vehicle,  8.97) (System Average Rate, 7.31) (Worst-Rate Vehicle, 4.32) (Second Worst-Rate Vehicle, 4.55) (System Rate Standard Deviation, 1.24)}; \addlegendentry{Proposed BGM-PA-MPM}
	
	\addplot[fill = color7] coordinates {(Highest-Rate Vehicle,  8.97) (System Average Rate, 7.23) (Worst-Rate Vehicle, 5.49) (Second Worst-Rate Vehicle, 5.61) (System Rate Standard Deviation, 1.02)}; \addlegendentry{Proposed BGM-PA-COMB}
	
	\end{axis}
	\end{tikzpicture}
	\caption{Data rate for $N = 130, L = 100$ and $K = 7$ with $J=3, N_1=100, N_2=90, N_3=80, \hat{N} = 70$}
	\label{f8}
	\vspace{-0.205cm}
\end{figure*}

In Fig. \ref{f7}, we show 5 different criteria to evaluate the performance of the approaches. We can observe that BGM-SA attains near-optimality as its performance is within $0.5\%$ of error. As we had presumed, BGM-PA-MIN exhibits an acceptable performance compared to all other variants, being surpassed only by BGM-PA-COMB in most cases. Because BGM-PA-COMB is based on BGM-PA-MIN and in addition employs statistical information of the group of vehicles, it can in general achieve superior performance under all the five criteria. However, under the criterion \textit{system average rate}, the best performance within the BGM-PA variants is attained by BGM-PA-AVE. This behavior results logical because BGM-PA-AVE considers---by definition---the average channel conditions. Therefore, if BGM-PA-COMB had not been introduced, we could have expected BGM-PA-MIN to perform best under the \textit{worst-rate vehicle} criterion, for the same reasons explained above. The variant BGM-PA-MPM, which is based on BGM-PA-MIN, can also attain acceptable performance under most of the criteria. On the other hand, BGM-PA-MAX and BGM-PA-IVAR are not capable of attaining good performance under \textit{worst-rate vehicle} and \textit{system rate standard deviation}. These two criteria would usually exhibit a favorable behavior when the method can provide fairness. Nevertheless, since BGM-PA-MAX is based on a greedy principle and BGM-PA-IVAR is by itself insufficient, both variants perform poorly.

Fig. \ref{f8} illustrates a setup similar to Fig. \ref{f7} but with a change in the number of vehicles at the intersection, namely $\hat{N} = 70$. Thus, the number of vehicles in the system is $N = 130$. We can observe that because of the increment of $\hat{N}$, the performance of all the approaches have changed. In some cases the performance improves whereas in others degradation can be identified. Notice that BGM-SA still attains near-optimality but with a comparatively increased error of $3\%$ in contrast to the previous case. However, some BGM-PA variants have undergone a considerable upturn. The reason why the performance of BGM-SA has suffered degradation, is essentially due to the increase of number of vehicles at the intersection. More specifically, this means that when the first cluster $\mathcal{V}^{(1)}$ is processed, the best subchannels will be selected for its $N_1 = 100$ vehicles. When the turn of $\mathcal{V}^{(2)}$ comes, there will be $\hat{N} = 70$ time subframes already in use, leaving only $N_1 - \hat{N} = 30$ available. Thus, the $N_2 - \hat{N} = 20$ unalloted vehicles of $\mathcal{V}^{(2)}$ must be accommodated in those 30 remaining subframes. Notice that the remaining free subframes may not necessarily have subchannels with high SINR for the vehicles in $\mathcal{V}^{(2)}$, as this was never enforced during the allocation of $\mathcal{V}^{(1)}$. If there were fewer vehicles at the intersection, e.g. $\hat{N} = 30$ as in Fig. \ref{f7}, BGM-SA would be able to achieve higher performance as more unused subframes would be available. On the other hand, we observe the opposite effect in BGM-PA. When the number of vehicles at the intersection $\hat{N}$ increases, its performance is boosted. The explanation to this outcome is that vehicles at the intersection are not grouped (this is done in order to prevent conflicts). Thus, there are $\hat{N} =70$ vehicles at the intersection and at most 30 lying outside the that area (prior to pre-grouping). And as we may infer, the main performance degradation source for BGM-PA is grouping due to the difficulty of representing channel conditions of a group with a single metric. Thus, since there are fewer groups of vehicles compared to the previous case, the performance is improved. If we had considered a larger number of vehicles at the intersection such as $\hat{N} = 95$ with $N_1 = N_2 = N_3 = 100$, the performance of both BGM-PA-MIN and BGM-PA-COMB would have been within $6\%$ of optimality.
\begin{figure}[!t]
	\centering
	\begin{tikzpicture}
	\begin{axis}[
	ymin = 1,
	ymax = 7,
	xmin = 0.09,
	xmax = 1,
	width = 9.2cm,
	height = 6.5cm,
	xlabel={$\hat{N} \slash Nj$},
	x label style={align=center, font=\footnotesize,},
	ylabel = {Rate [Mbits / s / subchannel]},
	y label style={at={(-0.06,0.5)}, text width = 4cm, align=center, font=\footnotesize,},
	ytick = {1, 2, 3, 4, 5, 6, 7, 8},
	enlarge y limits = {value = 0.4, upper},
	legend columns=2,
	legend pos = north east,
	legend style={font=\fontsize{7}{6}\selectfont, text width=2.75cm,text height=0.07cm,text depth=.ex, fill = none, }],
	]
	
	\addplot [color=color0, mark = star, mark options = {scale = 0.8, fill = blue}, line width = 0.8pt, style = dashed] coordinates 
	{
		(0.1, 6.92)
		(0.2, 6.87)
		(0.3, 6.85)
		(0.4, 6.82)
		(0.5, 6.80)
		(0.6, 6.79)
		(0.7, 6.70)
		(0.8, 6.64)
		(0.9, 6.46)
		(0.99, 6.20)
	}; \addlegendentry{Exhaustive Search}

	\addplot [color = black, mark = pentagon*, mark options = {scale = 1.5, fill = color1, solid}, line width = 1pt] coordinates 
	{
		(0.1, 6.89)
		(0.2, 6.86)
		(0.3, 6.83)
		(0.4, 6.77)
		(0.5, 6.70)
		(0.6, 6.60)
		(0.7, 6.52)
		(0.8, 6.31)
		(0.9, 6.06)
		(0.99, 5.68)  
	}; \addlegendentry{Proposed BGM-SA}
	
	\addplot[color=black, mark = triangle*, mark options = {scale = 1.5, fill = color2}, line width = 1pt] coordinates 
	{
		(0.1, 4.98)
		(0.2, 4.95)
		(0.3, 4.95)
		(0.4, 4.93)
		(0.5, 4.93)
		(0.6, 4.95)
		(0.7, 4.97)
		(0.8, 4.97)
		(0.9, 5.02)
		(0.99, 5.49)   
	}; \addlegendentry{Proposed BGM-PA-MIN}

	\addplot[color=black, mark = triangle*, mark options = {scale = 1.5, fill = color3}, line width = 1pt] coordinates 
	{
		(0.1, 1.47)
		(0.2, 1.51)
		(0.3, 1.54)
		(0.4, 1.55)
		(0.5, 1.65)
		(0.6, 1.69)
		(0.7, 1.82)
		(0.8, 1.94)
		(0.9, 2.22)
		(0.99, 3.61)   
	}; \addlegendentry{Proposed BGM-PA-MAX}

	\addplot[color=black, mark = triangle*, mark options = {scale = 1.5, fill = color4}, line width = 1pt] coordinates 
	{
		(0.1, 3.22)
		(0.2, 3.24)
		(0.3, 3.25)
		(0.4, 3.26)
		(0.5, 3.43)
		(0.6, 3.44)
		(0.7, 3.53)
		(0.8, 3.65)
		(0.9, 3.91)
		(0.99, 5.01)   
	}; \addlegendentry{Proposed BGM-PA-AVE}

	\addplot[color=black, mark = triangle*, mark options = {scale = 1.5, fill = color5}, line width = 1pt] coordinates 
	{
		(0.1, 1.41)
		(0.2, 1.45)
		(0.3, 1.46)
		(0.4, 1.50)
		(0.5, 1.58)
		(0.6, 1.59)
		(0.7, 1.69)
		(0.8, 1.83)
		(0.9, 2.08)
		(0.99, 3.27)   
	}; \addlegendentry{Proposed BGM-PA-IVAR}

	\addplot[color=black, mark = triangle*, mark options = {scale = 1.5, fill = color6}, line width = 1pt] coordinates 
	{
		(0.1, 3.65)
		(0.2, 3.66)
		(0.3, 3.72)
		(0.4, 3.73)
		(0.5, 3.76)
		(0.6, 3.84)
		(0.7, 3.90)
		(0.8, 4.00)
		(0.9, 4.17)
		(0.99, 5.15)   
	}; \addlegendentry{Proposed BGM-PA-MPM}

	\addplot[color=black, mark = triangle*, mark options = {scale = 1.5, fill = color7}, line width = 1pt] coordinates 
	{
		(0.1, 4.95)
		(0.2, 4.97)
		(0.3, 4.99)
		(0.4, 5.00)
		(0.5, 5.02)
		(0.6, 5.05)
		(0.7, 5.07)
		(0.8, 5.12)
		(0.9, 5.18)
		(0.99, 5.67)   
	}; \addlegendentry{Proposed BGM-PA-COMB}
	
	\end{axis}
	\end{tikzpicture}
	\caption{Worst-rate vehicle for $L = 100, K = 7$ with $J=4, N_1=100, N_2=100, N_3=100, N_4=100$ and varying $\hat{N}$.}
	\label{f9}
\end{figure}
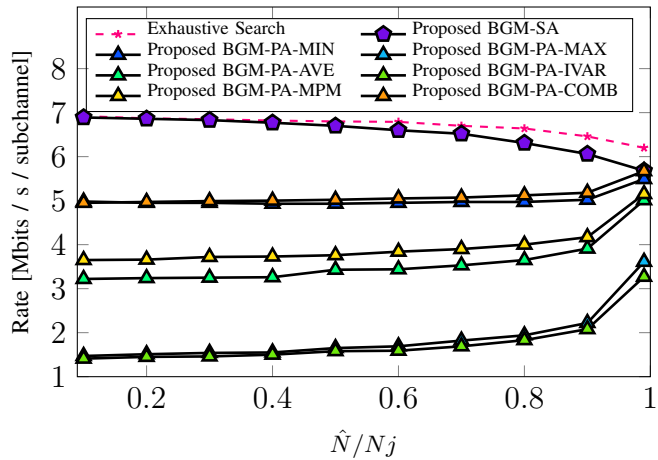
\begin{figure}[!t]
	\centering
	\begin{tikzpicture}[spy using outlines={circle, magnification=1.5, connect spies}]
	\begin{axis}[
	xmin = 5,
	xmax = 9,
	ymin = 0,
	ymax = 1,
	width = 9.2cm,
	height = 6.5cm,
	xlabel={Rate\textsubscript{x} [bits / s / Hz]},
	x label style={align=center, font=\footnotesize,},
	ylabel = {Pr\big(Rate~\textless~Rate\textsubscript{x}\big)},
	y label style={at={(-0.06,0.5)}, text width = 3.5cm, align=center, font=\footnotesize,},
	ytick = {0.1, 0.4, 0.7, 1.0},
	enlarge y limits = {value = 0.45, upper},
	legend columns=2,
	legend pos = north east,
	legend style={font=\fontsize{7}{6}\selectfont, text width=2.75cm,text height=0.07cm,text depth=.ex, fill = none, }
	]
	
		\addplot [color=color0, mark = star, mark options = {scale = 0.8, fill = blue}, line width = 0.8pt, style = dashed] table {CDF-ES-2.txt}; \addlegendentry{Exhaustive Search}
	
		\addplot [color = black, mark = pentagon*, mark options = {scale = 1.5, fill = color1, solid}, line width = 1pt] table {CDF-BGM-SA-2.txt}; \addlegendentry{Proposed BGM-SA}
	
		\addplot[color=black, mark = triangle*, mark options = {scale = 1.5, fill = color2}, line width = 1pt] table {CDF-BGM-PA-MIN-2.txt}; \addlegendentry{Proposed BGM-PA-MIN}
		
		\addplot[color=black, mark = triangle*, mark options = {scale = 1.5, fill = color3}, line width = 1pt] table {CDF-BGM-PA-MAX-2.txt}; \addlegendentry{Proposed BGM-PA-MAX}
		
		\addplot[color=black, mark = triangle*, mark options = {scale = 1.5, fill = color4}, line width = 1pt] table {CDF-BGM-PA-AVE-2.txt}; \addlegendentry{Proposed BGM-PA-AVE}
		
		\addplot[color=black, mark = triangle*, mark options = {scale = 1.5, fill = color5}, line width = 1pt] table {CDF-BGM-PA-IVAR-2.txt}; \addlegendentry{Proposed BGM-PA-IVAR}
		
		\addplot[color=black, mark = triangle*, mark options = {scale = 1.5, fill = color6}, line width = 1pt] table {CDF-BGM-PA-MPM-2.txt}; \addlegendentry{Proposed BGM-PA-MPM}
		
		\addplot[color=black, mark = triangle*, mark options = {scale = 1.5, fill = color7}, line width = 1pt] table {CDF-BGM-PA-COMB-2.txt}; \addlegendentry{Proposed BGM-PA-COMB}
	
	\end{axis}
	\end{tikzpicture}
	\caption{Cumulative distribution function (CDF) of rate values for $L = 100, K = 7$ with $J = 3, N_1 = 100, N_2 = 90, N_3 = 80$ and $\hat{N} = 50$.} 
	\label{f10}
	\vspace{-0.4cm}
\end{figure}
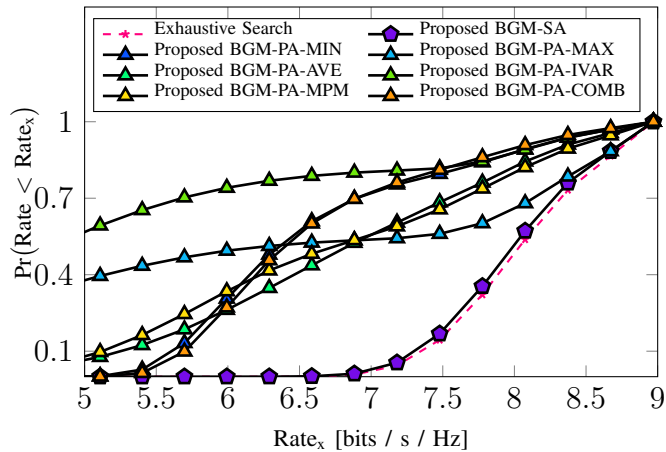

Fig. 9 shows the data rate experienced by the \textit{worst-rate vehicle}. In the abscissa, we vary the ratio $\hat{N} \slash N_j$ which represents the proportion of vehicles at the intersection to vehicles in each cluster. In this setup, we have considered that $N_1 = N_2 = N_3 = N_4 = 100$ and $J = 4$ clusters. For the reasons explained above, we expect that as the ratio $\hat{N} \slash N_j$ approaches unity the performance of BGM-SA will decrease whereas the performance of BGM-PA will increase. In our opinion, leveraging the \textit{worst-rate vehicle} is a most important criterion as it guarantees a minimum achievable rate for the least favored vehicle. Thus, judging from the results, we can say that the proposed BGM-SA, BGM-PA-MIN and BGM-PA-COMB are robust allocation schemes that are not prone to influence stemming from the diversity of possible scenarios.

Fig. 10 shows the cumulative distribution function (CDF) of the achievable rates. In this scenario, we have considered $J = 3$  clusters with $N_1 = 100, N_2 = 90, N_3 = 80$. Also, we have chosen $\hat{N} = 50$ as it is an intermediate value between the most and least favorable scenarios for BGM-SA. We observe that BGM-SA is similar in performance to exhaustive search, and is undoubtedly superior to all other approaches. We know, however, that such additional gain is achieved at the expense of higher complexity. We also observe that the second and third best schemes are BGM-PA-COMB and BGM-PA-MIN, respectively. Specifically, these two variants perform well in the low regime whereas they do not excel in the large regime. On th other hand, BGM-PA-MAX only performs well in the large regime. For this reason, BGM-PA-MPM---which uses both the \textit{MAX} and \textit{MIN} metrics---also performs acceptably right in the whole range. 

\section{Conclusion}
We have presented two resource allocation schemes for V2V broadcast communications. BGM-SA is based on successive matchings of weighted bipartite graphs whereas BGM-PA is capable of accomplishing the allocation---for all the clusters in the system---in a parallel fashion. We showed through simulations that BGM-SA can attain near-optimality with a complexity that increases proportionally to the number of clusters. On the other hand, BGM-PA has a lower complexity but achieves inferior performance. We also presented six different metrics to improve the matching performance of BGM-PA. Thus, the variants BGM-PA-COMB and BGM-PA-MIN are the most robust since they are not influenced by the system setup. In the allocation process, we always considered the enforcement of constraints in order to avoid intra-cluster allocation conflicts. A naive assumption of this work is that clusters can always be perfectly defined although in practice this might be complicated to guarantee.

\end{document}